\documentclass[aps,reprint,twocolumn,showpacs]{revtex4}
\usepackage{amsmath}
\usepackage{epsfig}
\usepackage{dcolumn}
\usepackage{bm}
\usepackage{amsfonts}
\usepackage{graphicx}
\usepackage{color}

\begin{document}

\title{Qubit dynamics at tunneling Fermi-edge singularity in \textit{a.c.}
response.}
\author{V.V. Ponomarenko$^{1,3}$ and I. A. Larkin$^{2,3}$}
\affiliation{$^1$Nonlinearity and Complexity Research Group, Aston University, Birmingham
B4 7ET, United Kingdom \\
$^2$Institute of Microelectronics Technology RAS, 142432 Chernogolovka,
Russia \\
$^3$Center of Physics, University of Minho, Campus Gualtar, 4710-057 Braga,
Portugal}
\date{\today }

\begin{abstract}
We consider tunneling of spinless electrons from a single-channel emitter
into an empty collector through an interacting resonant level of the quantum
dot. When all Coulomb screening of sudden charge variations of the dot
during the tunneling is realized by the emitter channel, the system is
described with an exactly solvable model of a dissipative qubit.
To study manifestations of the coherent
qubit dynamics in the collector \textit{a.c.} response we derive
solution to the corresponding Bloch equation for the model quantum
evolution in the presence of the oscillating  voltage  of frequency $%
\omega$ and calculate perturbatively the \textit{a.c.} response in the voltage amplitude.
We have shown that in a wide range of the model parameters the coherent qubit
dynamics results in the non-zero frequencies resonances in the
amplitudes dependence of the \textit{a.c.} harmonics
and in the jumps of the harmonics phase shifts across the resonances.
In the first order the \textit{a.c.} response is
directly related to the spectral decomposition of the corresponding transient current and
contains  only the first $\omega$ harmonic, whose amplitude exhibits resonance
at $\omega =\omega_I $, where $\omega_I$ is the qubit oscillation frequency.
In the second order we have obtained the $2 \omega$ harmonic of the a.c. response
with resonances in the frequency dependence of its amplitude at $\omega_I$, $%
\omega_I/2$ and zero frequency and also have found the frequency dependent shift
of the average steady current.

\end{abstract}

\maketitle

\section{Introduction}

The generic response of conduction electrons in a metal to the sudden
appearance of a local perturbation results in the Fermi-edge singularity
(FES) initially predicted in \cite{1,2} and also studied at finite
temperature \cite{oth} and more recently in the non-equilibrium Fermi
systems \cite{2muz,2a}. It was observed experimentally as a power-law
singularity in X-ray absorption spectra\cite{3,4}. Later, a possible
occurrence of the FES in transport of spinless electrons through a quantum
dot (QD) was considered \cite{5} in the regime when a localized QD level is
above the Fermi level of the collector in its proximity and the emitter is
filled up to the high energies. In the equivalent through the particle-hole
symmetry formulation realized in some experiments \cite{gei,6,7}, which we
follow in this work, the collector is effectively empty and the localized QD
level is close to the Fermi level of the emitter.  The Coulomb interaction
with the charge of the local level acts as a one-body scattering potential
for the electrons in the emitter. Then, in the perturbative approach
assuming a sufficiently small tunneling rate of the emitter, the separate
electron tunnelings from the emitter change the level occupation and
generate sudden changes of the scattering potential leading to the FES in
the I-V curves at the voltage threshold corresponding to the resonance.
Direct observation of these perturbative results in experiments, however, is
difficult because of the finite life time of electrons in the localized
state of the QD, and in many experiments \cite{gei,6,7,9} the FES's have
been identified simply by appearance of the threshold peaks in the I-V
dependence. According to the FES theory \cite{1,2} such a peak could occur
when the exchange effect of the Coulomb interaction in the tunneling channel
exceeds the Anderson orthogonality catastrophe effects in the screening
channels and, therefore, it signals the formation of an exciton
electron-hole pair in the tunneling channel at the QD. This pair can be
considered as a two-level system or qubit which undergoes dissipative
dynamics. In the absence of the collector tunneling and, if the Ohmic
dissipation produced by the emitter is weak enough, its dynamics are
characterized \cite{L,Sch} by the oscillating behavior of the level
occupation as a consequence of the qubit coherent dynamics, which is beyond
the perturbative description. The recent shot noise measurements \cite%
{8,noise} at the FES have raised a new interest \cite{novotny} to the qubit
dynamics, though direct realization of its coherency in the shot noise
should be further clarified.

Therefore, in this work we study the qubit dynamics and manifestation of its
coherency in the collector \textit{a.c.} response to a periodic time dependent
voltage.
We consider this in the same simplified, but still realistic system we used earlier \cite{epl}
to examine the transient tunneling current behavior.
In this system all sudden variations in charge of
the QD are effectively screened by a single tunneling channel of the
emitter. It can be realized, in particular, if the emitter is represented by
a single edge-state in the integer quantum Hall effect.
This system is described by a model permitting an exact solution
in the absence of the time dependent voltage. Making use of this solution it
has been demonstrated \cite{epl} that the FES in the tunneling current vs. the constant
bias voltage should be accompanied by oscillations of the time-dependent
transient tunneling current in a wide range of the model parameters.
Although the predicted oscillations could be the most direct evidence of the
qubit coherent dynamics, their experimental observation involves measurement
of the time dependent transient current averaged over its quantum
fluctuations, which is a challenging experimental task. Therefore in this
work we consider a more practical way for experimental observation of the
coherent qubit dynamics through the measurement of the \textit{a.c.}
tunneling into the collector. In order to explore this approach we derive
the general solution to this model in the form which distinguishes the
transient qubit dynamics dependent on the initial condition for the qubit
evolution from its steady behavior in the long time limit in the presence of
a periodic time dependent voltage. From this solution we find the qubit
Bloch vector time dependence in the steady regime as a perturbative series
in the \textit{a.c.} voltage amplitude and further use this expression to
demonstrate how the oscillatory behavior of the transient tunneling current
emerges in the frequency dependence of the parameters of the steady \textit{%
a.c.} response. In particular, we find that the oscillatory behavior of the
transient current results in the non-zero frequency resonances in the
amplitudes dependence of the \textit{a.c.} harmonics including the frequency
dependent shift of the average current and in the jumps of the \textit{a.c.}
harmonics phase shifts across the resonances. This confirms that the observation
of the \textit{a.c.} tunneling into the collector opens a realistic way for
the experimental demonstration of the coherent qubit dynamics and should be useful
for further identification of the FES in tunneling experiments. Note, that
being controlled by the tunneling into the empty collector, the \textit{a.c.}
response in this model remains independent of temperature.

The paper is organized as follows. In Sec. II we introduce the model and
formulate those conditions, which make it solvable through a standard
mapping onto the dissipative two-level system or qubit. In Sec. III we apply
the non-equilibrium Keldysh technique to derive Bloch equations
describing the dissipative evolution of the Bloch vector of the qubit
density matrix in the presence of a time-dependent voltage.

In Sec. IV their general solution is found, which in the case of a periodic
time dependent voltage permits us to describe the steady dynamics of the
qubit in the limit of the long time evolution and in the case of a constant
voltage reduces to the earlier developed description \cite{epl} of its
transient dynamics. Making use of this solution we introduce the main
characteristics of the transient dynamics including the Bloch vectors of the
qubit stationary states and their dependence on the experimentally
adjustable parameters of the setup, the decaying modes of the transient
qubit evolution, and connection between their corresponding amplitudes and
the qubit initial condition. The found expressions for these characteristics
are further used to establish the direct relation of the qubit transient
dynamics to its steady behavior.

In Sec. V we specify the steady behavior of the Bloch vector and the \textit{%
a.c.} tunneling response in the presence of a weakly oscillating voltage of
frequency $\omega$ perturbatively with respect to the voltage oscillation
amplitude and calculate the general term of the perturbative series. Next,
we analyze analytically and numerically the two lowest orders of this
expansion in details. In the first order of the perturbation expansion we
have only the first $\omega$ harmonic of the \textit{a.c.} response. Its
amplitude exhibits resonant behavior at $\omega =\omega_I $ for a wide range
of the model parameters, where $\omega_I$ is the frequency of the
oscillating transient current. In this order the \textit{a.c.} response is
directly related to the spectral decomposition of the transient current
produced by the specific initial disturbance of the qubit stationary state
by the applied voltage. This particular choice of the disturbance of the
stationary qubit state results in the suppression of the non-zero frequency
resonances of the \textit{a.c.} amplitude at the resonant QD level position.

In the second order of the Bloch vector expansion with respect to the
voltage amplitude we obtain its $2 \omega$ harmonic. It shows resonances in
the frequency dependence of the a.c. amplitude at $\omega_I$, $\omega_I/2$
and zero frequency. We also find that in this expansion order the frequency
dependent shift of the average steady current exhibits resonance at $%
\omega_I $ as a consequence of the transient current oscillations.

The results of the work are summarized in Conclusion, where we also compare
parameters of the system we consider with those realized in the recent experiments.

\section{Model}

In the system we consider below, the tunneling occurs from a single-channel
emitter into an empty collector through a single interacting resonant level
of the QD located between them. It is described with the Hamiltonian $%
\mathcal{H}=\mathcal{H}_{res}+\mathcal{H}_C$ consisting of the one-particle
Hamiltonian of resonant tunneling of spinless electrons and the Coulomb
interaction between instant charge variations of the dot and electrons in
the emitter. The resonant tunneling Hamiltonian takes the following form
\begin{equation}
\mathcal{H}_{res}=\epsilon _{0} d^+d+\sum_{a=e,c}\mathcal{H}_0[\psi_a]+
w_a(d^+\psi_a(0)+h.c.) \ ,  \label{hres}
\end{equation}
where the first term represents the resonant level of the dot, whose energy
is $\epsilon _{d}$. Electrons in the emitter (collector) are described with
the chiral Fermi fields $\psi_a(x),a=e(c)$, whose dynamics is governed by
the Hamiltonian $\mathcal{H}_0[\psi]=-i\!\! \int\! dx \psi^+(x) \partial_x
\psi(x) \ ( \hbar=1) $ with the Fermi level equal to zero or drawn to $%
-\infty$, respectively, and $w_a$ are the corresponding tunneling
amplitudes. The further application of the additional \textit{ac} voltage $%
-\delta V$ to the emitter can be accounted for in the Hamiltonian (\ref{hres}%
) with the time dependent bias of the resonant level energy $\epsilon
_{D}(t)=\epsilon _{d}+\delta V(t)$ in the case of the empty collector.

The Coulomb interaction in the Hamiltonian $\mathcal{H}$ is introduced as
\begin{equation}
\mathcal{H}_C=U_C \psi_e^+(0)\psi_e(0)(d^+d-1/2) \ .  \label{hc}
\end{equation}
Its strength parameter $U_C$ defines the scattering phase variation $\delta$
for the emitter electrons passing by the dot and therefore the screening
charge in the emitter produced by a sudden electron tunneling into the dot
is equal to $\Delta n=\delta/\pi \ \ (e=1)$ according to Friedel's sum rule.
Below we assume that the dot charge variations are completely screened by
the emitter tunneling channel and $\delta=-\pi$.

Next we implement bosonization and represent the emitter Fermi field as $%
\psi_e(x)=\sqrt{\frac{D}{2\pi} }\eta e^{i\phi(x)}$, where $\eta$ denotes an
auxiliary Majorana fermion and $D$ is the large Fermi energy of the emitter.
The chiral Bose field $\phi(x)$ satisfies $[\partial_x\phi(x),\phi(y)]=i2\pi%
\delta(x-y) $ and permits us to express
\begin{equation}
\mathcal{H}_0[\psi_e]=\int \frac{dx}{4 \pi} (\partial_x \phi)^2, \ \
\psi_e^+(0)\psi_e(0)=\frac{1}{2 \pi} \partial_x \phi(0) \ .  \label{hphi}
\end{equation}
Substituting these expressions into Eqs. (\ref{hres},\ref{hc}) we find the
alternative form for the Hamiltonian $\mathcal{H}$. By applying the unitary
transformation $\mathcal{U}=\exp[i\phi(0)(d^+d-1/2)]$ to this form we come
to the Hamiltonian of the dissipative two-level system or qubit:
\begin{eqnarray}
\mathcal{H}_{Q}(t)=\epsilon _{D}(t) d^+d+\mathcal{H}_0+
w_c(\psi^+_c(0)e^{i\phi(0)}d+h.c.)  \notag \\
+\Delta \eta (d- d^+)+ (\frac{U_C}{2\pi}-1) \partial_x \phi(0) (d^+d-\frac{1%
}{2}) \ ,  \label{hq} \\
\mathcal{H}_0=\mathcal{H}_0[\phi]+\mathcal{H}_0[\psi_c] \ ,  \notag
\end{eqnarray}
where $\Delta = \sqrt{\frac{D}{2\pi }}w_{e}$. This Hamiltonian is further
simplified. Since in the bosonization technique the relation \cite{schotte}
between the scattering phase and the Coulomb strength parameter is linear $%
\delta=-U_C/2$, the last term of the Hamiltonian on the right-hand side of
Eq. (\ref{hq}) vanishes and also the bosonic exponents in the third term can
be removed because the time dependent correlator of the collector electrons
is $<\psi_c(t)\psi^+_c(0)>=\delta(t)$.

\section{Bloch equations for the qubit evolution}

We use this Hamiltonian to describe the dissipative evolution of the qubit
density matrix $\rho _{a,b}(t)$, where $a,b=0,1$ denote the empty and filled
levels, respectively. In the absence of the tunneling into the collector at $%
w_{c}=0$, $\mathcal{H}_{Q}$ in Eq. (\ref{hq}) transforms through the
substitutions of $\eta (d-d^{+})=\sigma _{1}$ and $d^{+}d=(\sigma _{3}+1)/2$
( $\sigma _{1,3}$ are the corresponding Pauli matrices) into the Hamiltonian
$\mathcal{H}_{S}$ of a spin $1/2$ rotating in the magnetic field $\mathbf{h}%
(t)=(2\Delta , 0,\epsilon _{D}(t))^{T}$. Then the evolution equation follows
from
\begin{equation}
\partial _{t}\rho (t)=-i[\rho (t),\mathcal{H}_{S}(t)]\ .  \label{rhos}
\end{equation}%
To incorporate in it the dissipation effect due to tunneling into the empty
collector we apply the diagrammatic perturbative expansion of the S-matrix
defined by the Hamiltonian (\ref{hq}) in the tunneling amplitudes $w_{e,c}$
in the Keldysh technique. This permits us to integrate out the collector
Fermi field in the following way. At an arbitrary time $t$ each diagram
ascribes indexes $a(t_{+})$ and $b(t_{-})$ of the qubit states to the upper
and lower branches of the time-loop Keldysh contour. This corresponds to the
qubit state characterized by the $\rho _{a,b}(t)$ element of the density
matrix. The expansion in $w_{e}$ produces two-leg vertices in each line,
which change the line index into the opposite one. Their effect on the
density matrix evolution has been already included in Eq. (\ref{rhos}). In
addition, each line with index $1$ acquires two-leg diagonal vertices
produced by the electronic correlators $<\psi _{c}(t_{\alpha })\psi
_{c}^{+}(t_{\alpha }^{\prime }) >,\ \alpha =\pm $. They result in the
additional contribution to the density matrix variation: $\Delta \partial
_{t}\rho _{10}(t)=-\Gamma \rho _{10}(t),\ \Delta \partial _{t}\rho
_{01}(t)=-\Gamma \rho _{01}(t),\ \Delta \partial _{t}\rho _{11}(t)=-2\Gamma
\rho _{11}(t),\ \Gamma =w_{c}^{2}/2$. Then there are also vertical fermion
lines from the upper branch to the lower one due to the non-vanishing
correlator $<\psi _{c}(t_{-})\psi _{c}^{+}(t_{+}^{\prime })>$, which lead to
the variation $\Delta \partial _{t}\rho _{00}(t)=2\Gamma \rho _{11}(t)$.
Incorporating these additional terms into Eq. (\ref{rhos}) and making use of
the density matrix representation $\rho (t)=[1+\sum_{l}a_{l}(t)\sigma
_{l}]/2 $, we find the evolution equation for the Bloch vector $\mathbf{a}%
(t) $ as
\begin{equation}
\partial _{t}\mathbf{a}(t)=M(t)\cdot \mathbf{a}(t)+\mathbf{b}\ ,\ \mathbf{b}%
=[0,0,2\Gamma ]^{T}\ ,  \label{dadt}
\end{equation}%
where $M(t)$ stands for the matrix:
\begin{equation}
M(t)=\left(
\begin{array}{lll}
-\Gamma & -\epsilon _{D}(t) & 0 \\
\epsilon _{D}(t) & -\Gamma & -2\Delta \\
0 & 2\Delta & -2\Gamma%
\end{array}%
\right) \ .  \label{M}
\end{equation}%
It is divided into its stationary and time dependent parts: $M(t)=M_0+\Delta
M(t), \ \Delta M(t)=\delta V(t) \Lambda$, where the only non-zero matrix
elements of the $\Lambda$ matrix are $\Lambda_{1,2}=-\Lambda_{2,1}=-1$.

\section{Solution of the Bloch equations}

We find the general solution to Eq. (\ref{dadt}) describing the evolution of
the Bloch vector starting from its value $\mathbf{a}(0)$ at zero time in the
following form
\begin{eqnarray}
\mathbf{a}(t)=S(t,0)[\mathbf{a}(0)+M_0^{-1}\mathbf{b}]-M_0^{-1}\mathbf{b}
\notag \\
-\int_0^t dt^{\prime}S(t,t^{\prime})\Delta M(t^{\prime})M_0^{-1}\mathbf{b} \
.  \label{gat}
\end{eqnarray}%
Here the evolution operator $S(t,t^{\prime})=T_t\exp\{\int^t_{t^{\prime}}
d\tau M(\tau)\}$, where $T_t$ stands for the time-ordering, generalizes the
one $S_0(t)=\exp\{M_0 t\}$ for the time independent evolution. It can be
calculated perturbatively in $\Delta M$ in the interaction representation,
where
\begin{equation}
S(t,t_0)=S_0(t) T_t\left\{e^{\int^t_{t^{\prime}} d\tau \Delta
M_i(\tau)}\right\} S_0(-t_0)  \label{s}
\end{equation}
and $\Delta M_i(t)=S_0(-t_0)\Delta M(t)S_0(t)$.

The time independent evolution operator takes the following form through a
Laplace transformation:
\begin{equation}
S_0(t)=\int_{C}\frac{dz \ e^{zt}}{2\pi i}\left[ z-M_0\right] ^{-1}\ ,
\label{s0}
\end{equation}%
where the integration contour $C$ coincides with the imaginary axis shifting
to the right far enough to have all poles of the integral on its left side.
These poles are defined by inversion of the matrix $[z-M_0]$ and are equal
to three roots of its determinant $\det \left[ z-M_0\right] \equiv P(z)$,
which is
\begin{equation}
P(z)=x^{3}+\Gamma x^{2}+(4\Delta ^{2}+\epsilon _{d}^{2})x+\Gamma \epsilon
_{d}^{2} \ ,\ x=z+\Gamma \ .  \label{p3}
\end{equation}
From their explicit expressions below we conclude that all roots $z_{l},\
l=\left\{ 0,1,2\right\} $ have their real parts negative and, in the absence
of the \textit{ac} voltage, the evolution of the Bloch vector converges to
the second term on the right-hand side of Eq. (\ref{gat}).

\subsection{Stationary state of the qubit}

Therefore, in the absence of the \textit{ac} voltage the stationary state of
the qubit is characterized by the Bloch vector:
\begin{equation}
\mathbf{a}(\infty )=-M_0^{-1}\mathbf{b}=\frac{[2\epsilon _{d}\Delta
,-2\Delta \Gamma ,(\epsilon _{d}^{2}+\Gamma ^{2})]^{T}}{\left( \epsilon
_{d}^{2}+\Gamma ^{2}+2\Delta ^{2}\right) }\ .  \label{ainfty}
\end{equation}%
In general, an instant tunneling current $I(t)$ into the empty collector
directly measures the diagonal matrix element of the qubit density matrix
\cite{us} through their relation
\begin{equation}
I(t)=2\Gamma \rho _{11}(t)=\Gamma \lbrack 1-a_{3}(t)]  \label{I-t}
\end{equation}%
It gives us the stationary tunneling current as $I_{0}=2\Gamma
\Delta^{2}/(2\Delta ^{2}+\Gamma ^{2}+\epsilon _{d}^{2})$. At $\Gamma \gg
\Delta $ this expression coincides with the perturbative results of \cite%
{5,lar}. Another important characteristic is the qubit entanglement entropy $%
S_e=-\mbox{tr}\{\rho \ln \rho\}$, which is just a function of the Bloch
vector length:
\begin{equation}
S_e=\ln2-\ln(1-a^2)/2-\frac{a}{2} \ln\left[\frac{1+a}{1-a}\right] \ ,
\label{Se}
\end{equation}%
where $a=|\mathbf{a}|$. The length of the stationary Bloch vector in Eq. (%
\ref{ainfty}) is $a(\infty )=\sqrt{1-(I_0/\Gamma)^2}$. Therefore,
measurement of the tunneling current gives us also the entropy of the
stationary state of the qubit. This entropy changes from zero for the qubit
pure state of empty QD with $a=1$ far from the resonance to its entanglement
maximum approaching $\ln 2$ at the resonance with $a \approx 0$, if $\Gamma
\ll \Delta$.

\subsection{Qubit transient dynamics}

The first term on the right-hand side of Eq. (\ref{gat}) describes the
transient evolution of the Bloch vector, which is caused by the deviation of
the initial vector $\mathbf{a}(0)$ at zero time from its stationary state.
In the absence of the \textit{ac} voltage we find this vector $\mathbf{a}(t)$
at positive time through substitution of $S_0$ from Eq. (\ref{s0}) into Eq. (%
\ref{gat}) and closing the contour $C$ in the left half-plane as follows
\begin{equation}
\mathbf{a}(t)=\mathbf{a}(\infty )+\sum\limits_{l=0}^{2}\mathbf{r}_{l}\cdot
\exp [z_{l}t]  \label{at3}
\end{equation}
where the residues $\mathbf{r}_{l,\alpha}=\mathbf{l}_{R,\alpha} (\mathbf{l}%
_L|\mathbf{a}(0)-\mathbf{a}(\infty ))$ are expressed in terms of the right
and left normalized eigenvectors, $\mathbf{l}_R$ and $\mathbf{l}_L$,
corresponding to the $z_{l}$ eigenvalue of $M_{0}$. The three roots $z_{l}$
of $P(z)$ in Eq. (\ref{p3}) are defined by its coefficients through the two
parameters:
\begin{equation}
Q=12\Delta ^{2}-\Gamma^2 +3\epsilon _{d}^{2} ,\ \ \ R=\left( 18\Delta
^{2}-9\epsilon _{d}^{2}-\Gamma^2 \right)\Gamma ,  \label{QR}
\end{equation}
in the following way
\begin{align}
z_0&=-G_0= \gamma_1-\frac{4}{3}\Gamma ,\ \gamma _{1}=\frac{1}{3}(S+T)
\label{z-roots} \\
z_{1,2}&=-G_1 \pm i \omega_I ,\ \omega_I= \frac{\sqrt{3}}{6}(S-T) ,\ G_1=%
\frac{4}{3}\Gamma +\frac{\gamma_{1}}{2} \ ,  \notag
\end{align}
where
\begin{equation}
S=\left( R+\sqrt{Q^{3}+R^{2}}\right)^{1/3}\text{ and } T=-\frac{Q}{S} \ .
\label{ST}
\end{equation}
Here the function $\ Z^{1/3}$ of the complex variable $Z$ is determined in
the conventional way with the cut $Z\in \{-\infty ,0\}$. At the resonance
position of the level energy $(\epsilon _{d}=0)$ these eigenvectors and the
eigenvalues take particular simple form, since the evolution of the first
component of the Bloch vector in Eq. (\ref{dadt}) becomes independent of the
two other components.

\begin{figure}[t]
\centering \includegraphics[width=8cm]{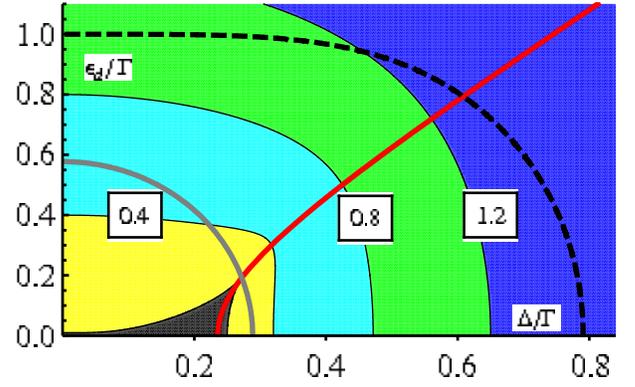}
\caption{ Contour plot of the positive imaginary part of the dimensionless
root Im$[y_{1}]/\Gamma=\frac{\protect\sqrt{3}}{2\Gamma}(S-T)$. The black
area corresponds to the region where all three roots are real. The red
(light gray) line corresponds to $R=0$ and the gray line to $Q=0$. The black
dashed curve shows $Im[z_1]=-Re[z_1] $.}
\label{fig:Im}
\end{figure}

If the discriminant is positive: $Q^{3}+R^{2}>0$, $S$ and $T$ are real
positive and negative, respectively. Therefore, the root $z_0$ is real and
the two others $z_{1,2}$ are complex conjugates of each other. In the case
of $Q^{3}+R^{2}<0$, $S$ and $T$ are also complex conjugate. Hence, all three
roots are real negative. In this case the oscillatory behavior of the Bloch
vector in Eq. (\ref{at3}) does not occur. This parametric area of triangular
form is depicted as black in Fig.\ref{fig:Im}. Its three vertices have
coordinates (0,0), (1/4,0) and ($\sqrt{2/27},\sqrt{1/27}$). In the
parametric area above the red line $R=0$ in Fig.\ref{fig:Im} both $R$ and $%
\gamma_1$ are negative. The $z_0$ non-oscillating mode of the Bloch vector
in Eq. (\ref{at3}) decays quicker than the oscillating modes and the
transient current is an infinitely oscillating function of time. Below this
line $\gamma_1$ is positive, the amplitude of the oscillations vanishes more
quickly than the first term, and the additional condition \cite{epl} on the
initial state of the QD should be fulfilled to observe oscillations of the
transient current. Below we consider manifestations of these oscillations in
the frequency dependent current response to the applied \textit{a.c.}
voltage of a small amplitude.

\section{Current response to the time dependent voltage}

The change in the Bloch vector evolution by the applied \textit{a.c.}
voltage is described in Eq. (\ref{gat}) by the first and third term on its
right-hand side with the evolution operator defined by Eq.(\ref{s}) with $%
\Delta M(t^{\prime})=v \cos(\omega t^{\prime})\Lambda$. Below we will be
interested in the steady regime of the evolution, when $\Gamma t \gg 1$ and
the starting time of the evolution can be drawn in Eq. (\ref{gat}) from zero
to $-\infty$. Then, contrary to the transient qubit dynamics, only the third
term contributes. The evolution operator in this term can be calculated
through expansion of the time-ordered exponent in Eq. (\ref{s}) with respect
to $\Delta M_i$ and we find the Bloch vector variation by the ac voltage
application as a perturbative series in $v$: $\Delta \mathbf{a}(t)=\sum
_{n=1}\Delta_n\mathbf{a}(t) $. The general term of this series $\Delta_n
\mathbf{a}(t)$ contains multiple $m\omega$ frequency components, where $m$
is of the same parity as $n$ and runs from zero to $n$. Below we calculate
explicitly the first two terms $\Delta_1 \mathbf{a}(t)$ and $\Delta_2
\mathbf{a}(t)$ of this series.
\begin{figure}[b]
\centering \includegraphics[width=8cm]{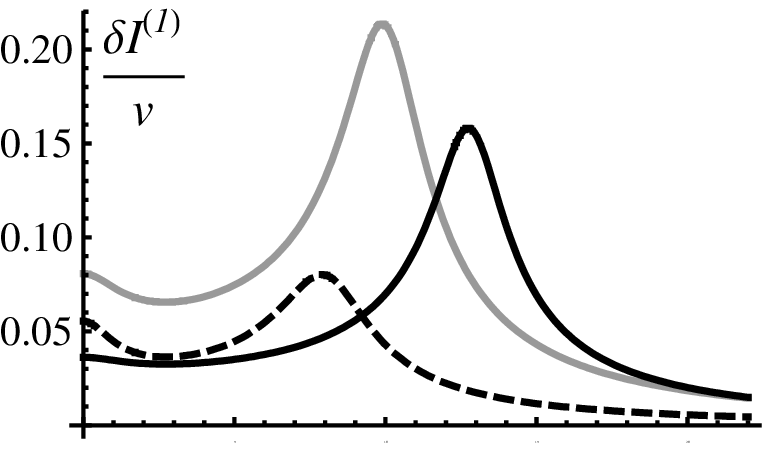} \centering %
\includegraphics[width=8cm]{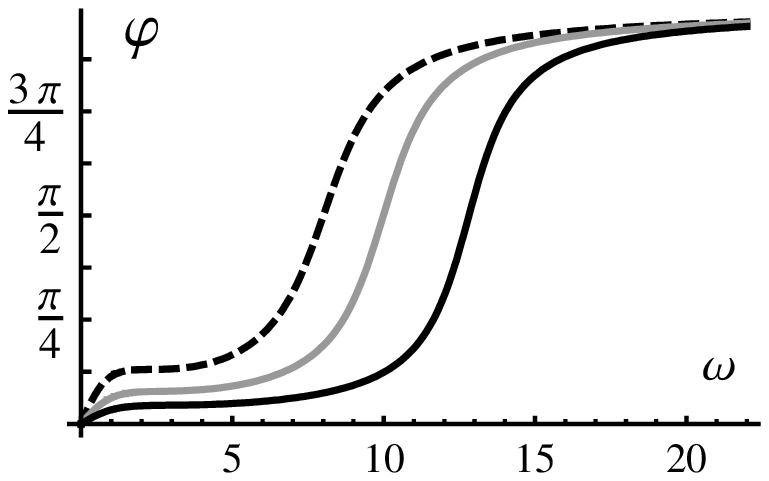}
\caption{Normalized amplitude (upper panel) and phase shift (lower panel) of
the first harmonic of the \textit{ac} current $\protect\delta I^{(1)}/v$
defined by Eqs. \protect\ref{phi}, \protect\ref{delta11} when $\Gamma =1$, $%
\Delta =4 $. The black lines correspond to the resonant level energy $%
\protect\epsilon_d =-10$, the gray lines to $\protect\epsilon_d =-6$, and the
black dashed lines to $\protect\epsilon_d =-1$ .}
\label{fig:I01}
\end{figure}

\subsection{Linear current response}

In the first order of the perturbation expansion in $v$ we substitute the $%
S_{0}$ operator from Eq. (\ref{s0}) for the evolution operator in the last
term on the right-hand side of Eq. (\ref{gat}). Choosing the contour $C$ in
Eq. (\ref{s0}) to go from $(-0-i\infty )$ to $(-0+i\infty )$ and performing
integration over $t^{\prime }$ in the last term, we further close the
contour in the right half-plane and express the result through the poles
contributions as
\begin{eqnarray}
\frac{\Delta _{1}\mathbf{a}(t)}{v} &=&Re\frac{e^{i\omega t}}{i\omega -M_{0}}%
\Lambda \mathbf{a}(\infty )=  \label{delta1} \\
&&Re\sum\limits_{l=0}^{2}\mathbf{l}_{R}\frac{e^{i\omega t}}{i\omega -z_{l}}(%
\mathbf{l}_{L}|\Lambda \mathbf{a}(\infty ))\ .  \notag
\end{eqnarray}%
From comparison of the last part of Eq. (\ref{delta1}) with Eq. (\ref{at3})
we conclude that in the $v$ linear response of the \textit{a.c.} Bloch
vector response $\Delta _{1}\mathbf{a}(t)$ coincides at times $t=%
\mbox{integer}\times 2\pi /\omega $ and $t=(\mbox{integer}-1/4)\times 2\pi
/\omega $ with the real and imaginary parts of the spectral decomposition of
the transient Bloch vector, originating from its initial state $\mathbf{a}%
(0)-\mathbf{a}(\infty )=\Lambda \mathbf{a}(\infty )$. The spectral
decomposition of the corresponding transient current $I_{tr}(t)$ can be found from the
oscillating current $\Delta _{1}I(t)=-\Gamma \Delta _{1}a_{3}(t)=\delta
I^{(1)}\cos (\omega t-\varphi )$ in the following way
\begin{equation}
\int_{0}^{\infty }dte^{-i\omega t}[I_{tr}(t)-I_{0}]=\delta _{1}I\cdot
e^{-i\varphi }\ ,  \label{phi}
\end{equation}%
where $\delta I^{(1)}$ and $\varphi $ are the amplitude and the phase shift
of $\Delta _{1}I(t))$. The oscillating behavior of the transient current defined by Eq. (%
\ref{at3}) emerges as a resonance in the frequency dependence of the
amplitude $\delta I^{(1)}$ and as an abrupt change of the phase $\varphi $
of the first $\omega $ harmonic of the oscillating current at $\omega
=\omega _{I}$ if the coefficient $\mathbf{r}_{1,3}$ at the resonant term in
Eq. (\ref{delta1}) is non zero.

By inversion of the matrix denominator in the first part of Eq. (\ref{delta1}%
) we find the oscillating current as follows
\begin{equation}
\Delta _{1}I(t)=-2I_{0}vRe[\epsilon _{d}(i\omega +2\Gamma )e^{i\omega
t}/P(i\omega )]  \label{delta11} \\
\end{equation}%
The amplitude of its oscillations $\delta I^{(1)}(\omega,\epsilon _{d})$ and the phase shift
$\varphi (\omega,\epsilon _{d})$ are depicted in Fig. \ref{fig:I01} for the several
choices of the parameters. Here we have defined $\varphi $ so that
it is positive at $\epsilon _{d}<0$ and $\omega >0$ and zero at $\omega=0$.
Then it expands as $\varphi (\omega, \epsilon _{d})=-\varphi (-\omega, \epsilon _{d})$ to negative $\omega$
and through  $\varphi (\omega, \epsilon _{d})=- \pi \,  \mbox{sgn}( \omega) +\varphi (\omega, -\epsilon _{d})$
for $ \epsilon _{d}>0$.  The
amplitude $\delta _{1}I(\omega,\epsilon _{d})$ is even function of both $\omega$ and $\epsilon_{d},$ namely,
$\delta _{1}I(\omega,\epsilon _{d})=\delta _{1}I(\pm\omega, -\epsilon_{d}).$
The coefficient $\mathbf{r}_{1,3}$ characterizing the strength of the $%
\delta I^{(1)}$ resonance follows from Eq. (\ref{delta11}) as
\begin{equation}
\mathbf{r}_{1,3}=-\frac{v I_{0}}{\Gamma^2 }\frac{\epsilon _{d}(2\Gamma
/3+\gamma _{1}/2+i\omega _{I})}{\omega _{I}(3\gamma _{1}-2i\omega _{I})}
\label{r1}
\end{equation}%
It vanishes at the resonance $\epsilon _{d}=0 $, where the deviation of the
Bloch vector initial condition from its stationary value $v\Lambda \mathbf{a}%
(\infty ) = v I_0/(\Gamma \Delta) [-\Gamma,\epsilon _{d}, 0]^T $ does not
produce the transient current.

At small $\Gamma \ll \sqrt{\epsilon _{d}^{2}+4\Delta ^{2}}$ the $\delta
I^{(1)}$ resonance in Fig. \ref{fig:I01} has Lorentzian shape centered at $%
\omega _{I}\approx \sqrt{\epsilon _{d}^{2}+4\Delta ^{2}}$ with the width $%
G_{1}\approx \Gamma (1+2\Delta ^{2}/\omega _{I}^{2}).$ Since the poles $z_{1,2}$
are close to the imaginary axis, if Re[$z_{1,2}]\ll \omega _{I}$, the phase
shift at $\omega =\omega_{I}$ is $\varphi \approx {\small \mp }\pi /2$.
\begin{figure}[t]
\centering \includegraphics[width=8cm]{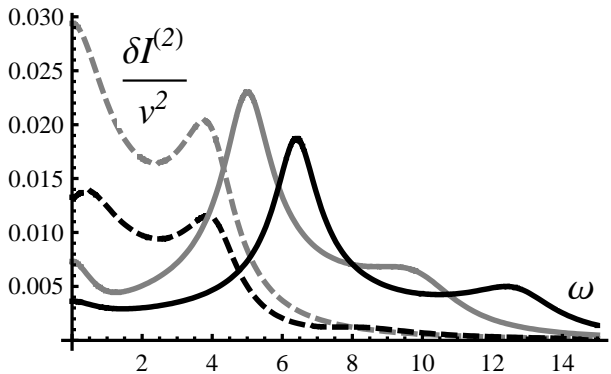}
\caption{Normalized amplitude of the second harmonic of the \textit{ac}
current $\protect\delta I^{(2)}/v^2$ from Eq. \protect\ref{delta22} when $%
\Gamma =1$, $\Delta =4 $. The black line corresponds to the resonant level
energy $\protect\epsilon_d =10$, the gray line to $\protect\epsilon_d =6$,
and the black dashed line to $\protect\epsilon_d =2, $ and the gray dashed
line to $\protect\epsilon_d =0$ .}
\label{fig:I22}
\end{figure}
\begin{figure}[b]
\centering \includegraphics[width=8cm]{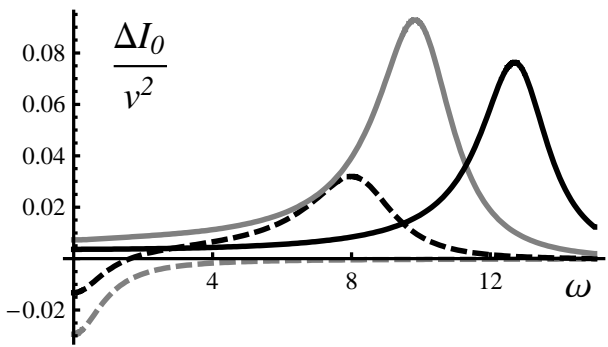}
\caption{The frequency dependent variation of the average current $\Delta
I_0/v^2$ from Eq. \protect\ref{delta220} when $\Gamma =1$, $\Delta =4 $. The
black line corresponds to the resonant level energy $\protect\epsilon_d =10$%
, the gray line to $\protect\epsilon_d =6$, and the black dashed line to $%
\protect\epsilon_d =2, $ and the gray dashed line to $\protect\epsilon_d =0$
.}
\label{fig:I02}
\end{figure}

\subsection{Second-order current response}

Substitution of the first $\Delta M_{I}$ order expansion of the expression
for the evolution operator $S(t,t_{0})$ in Eq. (\ref{s}) into the last term
on the right-hand side of Eq. (\ref{gat}) gives us
\begin{eqnarray}
\Delta _{2}\mathbf{a}(t) &=&  \label{delta21} \\
\int_{-\infty }^{t}\!\!\!\!\!\!d\tau \int_{-\infty }^{\tau }\!\!\!dt^{\prime
}S_{0}(t-\tau ) &\Delta &M(\tau )S_{0}(\tau -t^{\prime })\Delta M(t^{\prime
})\mathbf{a}(\infty )\ .  \notag
\end{eqnarray}%
Making use of the representation of $S_{0}$ in Eq. (\ref{s0}) and performing
successive integration over $t^{\prime }$ and $\tau $ we find $\Delta _{2}%
\mathbf{a}(t)$ in Eq. (\ref{delta21}) after closing both contours $C$ of the
Laplace transformation in the right half-plane as follows
\begin{eqnarray}
\Delta _{2}\mathbf{a}(t) &=&\frac{v^{2}}{2}Re\left\{ \frac{e^{i2\omega t}}{%
i2\omega -M_{0}}\Lambda \frac{1}{i\omega -M_{0}}\Lambda \mathbf{a}(\infty
)\right\}  \notag \\
&&-\frac{v^{2}}{2M_{0}}\Lambda Re\left\{ \frac{1}{i\omega -M_{0}}\Lambda
\mathbf{a}(\infty )\right\} \ .  \label{delta22}
\end{eqnarray}%
The first term here describes the Bloch vector variation resulting in the
second $\omega $ harmonic of the oscillating current $\Delta _{2}I(t)$,
whose amplitude $\delta I^{(2)}(\omega )$ as depicted in Fig. \ref{fig:I22}
exhibits two resonances at $\omega =\omega _{I}/2$ and $\omega =\omega _{I}$
if $\Gamma $ small enough. The second term contributes to the frequency
dependent deviation $\Delta I_{0}$ due the \textit{a.c.} voltage of the
average steady current from the stationary current $I_{0}$. From comparison
of this term with the right side of Eq. (\ref{delta1}) we express this
current shift through the first order variation of the Bloch vector:
\begin{equation}
\Delta I_{0}=-\frac{vI_{0}}{2\Gamma \Delta }(\mathbf{x}|\Delta _{1}\mathbf{a}%
(0)),  \label{deltaI0}
\end{equation}%
where $\mathbf{x}=[\Gamma ,\epsilon _{d},0]^{T}$. Therefore, it combines the
spectral decompositions of the transient dynamics of the first two
components of the Bloch vector. We use its explicit form:
\begin{equation}
\Delta I_{0}=-\frac{v^{2}I_{0}^{2}}{2\Gamma \Delta ^{2}}Re\left[ \frac{%
P_{1}(i\omega )}{P(i\omega )}\right] ,  \label{delta220}
\end{equation}%
where
\begin{eqnarray}
P_{1}(z) &=&\left( \Gamma ^{2}-\epsilon _{d}^{2}\right) x^{2}+\left( \Gamma
^{3}-3\Gamma \epsilon _{d}^{2}\right) x+2\Gamma ^{2}\left( 2\Delta
^{2}-\epsilon _{d}^{2}\right)  \notag \\
\ \ \ x &=&z+\Gamma
\end{eqnarray}%
to demonstrate in Fig. \ref{fig:I02} that at finite $\epsilon _{d}\neq 0$ it
exhibits a single resonance at $\omega =\omega _{I}$ with the width $%
G_{1}\approx \Gamma (1+2\Delta ^{2}/\omega _{I}^{2})$. It transforms into
the negative Lorentz function $\Delta I_{0}/v^{2}=-I_{0}^{2}\Gamma /(\omega
^{2}+\Gamma ^{2})/(2\Delta ^{2})$ at the resonant level position as a
consequence of the pure exponential decay dynamics of the first Bloch vector
component.

\section{Conclusion}

The tunneling of spinless electrons through an interacting resonant level of
a QD into an empty collector has been studied in the especially simple, but
realistic system, in which all sudden variations in charge of the QD are
effectively screened by a single tunneling channel of the emitter. As a
result the time evolution of the system in question has been reduced to the
dynamics of a dissipative two-level model. Its off-diagonal coupling
parameter $\Delta $ is equal to the bare emitter tunneling rate $\Gamma _{e}$
renormalized by the large factor $\sqrt{D/(\pi\Gamma_e)} ,$ whereas the
dumping parameter $\Gamma $ coincides with the tunneling rate of the
collector.
For a constant bias voltage the short time transient behavior of this model was studied
in \cite{epl} and it has been shown that the FES in the tunneling current dependence on
the voltage should be accompanied by oscillations of the time-dependent
transient tunneling current in the wide range of the model parameters. In
particular, they occur if the emitter tunneling coupling $\Delta $ or the
absolute value of the resonant level energy $|\epsilon _{d}|$ are large
enough in comparison with the collector tunneling rate $\Gamma $ and either $%
\Delta >\Gamma /4$ or $\epsilon _{d}^{2}>\Gamma ^{2}/27$ holds.

Although these
oscillations should confirm the emergence of
the qubit composed of an electron-hole pair at the QD and its coherent dynamics
in agreement with the FES theory it could be difficult to observe them directly,
since this involves measurement
of the time dependent transient current averaged over its quantum
fluctuations.
Therefore, in this work we have studied manifestations of the coherent
qubit dynamics in the collector \textit{a.c.} response to a periodic time dependent
voltage of the frequency $\omega$.

We have calculated the steady long time dependence of the qubit
Bloch vector and the tunneling current  perturbatively in any order
in the \textit{a.c.} voltage amplitude and further have used this expansion to
describe  the frequency dependence of the parameters of the steady \textit{%
a.c.} response. In particular, we have found that the oscillatory behavior of the
transient current results in the non-zero frequency resonances in the
amplitudes dependence of the \textit{a.c.} harmonics
and in the jumps of the \textit{a.c.}
harmonics phase shifts across the resonances. In the first order of this expansion
only the first $\omega$ harmonic arises in the \textit{a.c.} response. Its
amplitude exhibits resonant behavior at $\omega =\omega_I $ for a wide range
of the model parameters, where $\omega_I$ is the frequency of the
oscillating transient current. In this order the \textit{a.c.} response is
directly related to the spectral decomposition of the transient current
produced by the specific initial disturbance of the qubit stationary state
by the applied voltage.
In the higher orders of the \textit{a.c.} expansion also the higher harmonics emerge
and the corresponding resonances occur at the fractions of $\omega_I$.
The found results confirm that the observation
of the \textit{a.c.} tunneling into the collector opens a realistic way for
the experimental demonstration of the coherent qubit dynamics and should be useful
for further identification of the FES in tunneling experiments.

We have performed our calculations in dimensionless units with $\hbar =1$
and $e=1,$ therefore, the unit of the \textit{a.c.} admittance in Fig. \ref%
{fig:I01} is~$e^{2}/\hbar \approx 2.43\cdot 10^{-4}$S. In the experiments
\cite{lar,lar1} the collector tunneling rate is $\Gamma \approx 0.1 \, meV$ and
the coupling parameter $\Delta \approx 0.016 \, meV$. This corresponds to the
stationary current equal to $I_{0}\approx 1.2 \, nA$ at $\epsilon _{d}=0$. To
observe the regime of the induced oscillations shown in Fig. \ref{fig:I01}
one can increase the collector barrier width to obtain the heterostructure
with $\Delta =4\Gamma $. Then, with $\Delta =0.016 \, meV$ and $\Gamma \approx
4\mu eV$, the stationary current at the resonant level position is $%
I_{0}=0.94 \,nA$. The unit of frequency $\omega $ in Figs. \ref{fig:I01}-\ref%
{fig:I02} for this value of $\Gamma $ is $6.08\cdot10^{9}s^{-1}$. With the
amplitude of the oscillating voltage equal to $v=1\mu V$ we find that the
peak of the \textit{a.c.} amplitude of its first harmonic is $\delta
I_{mx}^{(1)}=53\,pA$. It occurs at $\nu _{I}=\omega _{I}/(2\pi)=9.6\,GHz$
for $\epsilon _{d}^{{}}=-0.024\,meV$ (gray curves in Fig. \ref{fig:I01}). At
the same parameters $v$ and $\epsilon _{d}$ the resonance of the \textit{a.c.%
} second harmonic (gray curve in Fig. \ref{fig:I22}) takes place at $\nu
_{I}/2=4.8\,GHz$ and its amplitude is $\delta I_{mx}^{(2)}=1.4\,pA$.

\section{Acknowledgment}

The work was supported by the Foundation for Science and Technology of
Portugal and by the European Union Seventh Framework Programme
(FP7/2007-2013) under grant agreement n$^{\mathrm{o}}$ PCOFUND-GA-2009-
246542 and Research Fellowship SFRH/BI/52154/2013. It was also funded (V.P.)
by a Leverhulme Trust Research Project Grant RPG-2016-044.

\end{document}